\documentstyle[aps,multicol,prb,graphics,times]{revtex}
\newcommand{\bfk}{{\bf k}}
\newcommand{\bfr}{{\bf r}}
\newcommand{\bfa}{{\bf a}}
\newcommand{\bfb}{{\bf b}}
\newcommand{\bfA}{{\bf A}}
\newcommand{\bfB}{{\bf B}}
\newcommand{\bfx}{{\bf x}}
\newcommand{\bfy}{{\bf y}}
\begin{document}
\draft
\title{Anomalous dynamic response in the two-dimensional lattice
  Coulomb gas model: Effects of pinning
  }
\author {Beom Jun Kim and  Petter Minnhagen}
\address {Department of Theoretical Physics,
Ume{\aa} University,
901 87 Ume{\aa}, Sweden}
\preprint{\today}
\maketitle
\begin{abstract}
It is demonstrated through Monte Carlo simulations that the 
one component lattice Coulomb gas model in two dimensions
under certain conditions display features of an anomalous dynamic response.
We suggest that pinning, which can either
be due to the underlying discrete
lattice or induced by disorder,
is an essential ingredient behind this
anomalous behavior. 
The results are discussed in relation to other situations 
where this response type appears, in particular the two components neutral
Coulomb gas below the Kosterlitz-Thouless transition, as well as 
in relation to
other findings from theory, simulations, and experiments on superconductors.     
\end{abstract}


\begin{multicols}{2}
\section{Introduction} \label{sec:intro}
Two-dimensional (2D) vortex physics is strongly reflected in the properties of
systems related to 2D superfluids like Josephson junction
arrays,~\cite{minnhagen-rev,newrock} superconducting films, and $^4$He
films,~\cite{minnhagen-rev} as well as high-$T_c$
superconductors.~\cite{minnhagen-rev,minnhagen-rev-highTc} The vortices are,
e.g.,  responsible for the well-known Kosterlitz-Thouless (KT)
transition~\cite{minnhagen-rev,kosterlitz,kosterlitz-rg} and there is a fairly
good understanding of the thermodynamic properties related to 2D vortex
physics. However, the dynamical aspects,  which can be probed by the complex
impedance and the flux noise measurements in
superconductors,~\cite{minnhagen-rev,minnhagen-rev-highTc,beom-noise} and by
the torsional oscillator period shift in $^4$He-films,~\cite{bishop} are much
less well understood.  The first attempt to describe 
the dynamic vortex response phenomenologically was given by 
Ambegaokar-Halperin-Nelson-Siggia
(AHNS) in Ref.~\onlinecite{ahns}. However, it was later discovered
that the Minnhagen
phenomenology (MP) in Ref.~\onlinecite{minnhagen-rev} based on essentially the same
ingredients as AHNS, described experiments and simulations
better.~\cite{wallin,rogers,theron,houlrik,holmlund,jonsson1,jonsson2,beom-big}
A particularly interesting aspect in the MP is that it 
reflects an anomaly in the 2D vortex response,~\cite{theron} which, for example,
in a superconductor takes the form of a logarithmic divergence of
the complex conductivity:  $\sigma(\omega)\propto -\ln \omega$ for
small frequency $\omega$.  There still remain open questions on 
what ingredients can cause this anomaly and give rise to the anomalous
response form. 

There has been a number of
attempts to obtain the MP form with other phenomenological approaches
as well as somewhat formally more rigorous ones.~\cite{beck,korshunov,capezalli,bormann,bowley1,bowley}
One possibility is that it is an intrinsic property of the 2D vortex
system which is linked to the long range logarithmic interaction, as
presumed by the original motivation for the MP response
form.~\cite{minnhagen-rev} Another possibility is that it is caused by the coupling
between the perpendicular currents described by the vortices and the
longitudinal currents (often referred to as the spin-wave
part).~\cite{beck,korshunov} In the present paper we suggest
that there is yet another possibility; the MP type
response can also arise from pinning influencing the vortex system.

The approach to this problem, which we subscribe to here,
is to systematically investigate various models by
aid of simulations. Vortices can be described as 2D Coulomb gas
particles, in the sense that a vortex with vorticity $\pm 1$
corresponds to a Coulomb gas particle with charge $\pm 1$ (See, e.g., 
Ref.~\onlinecite{minnhagen-rev} for details on the mapping between the
two models).
Since the dynamics of the vortices can to a good approximation
be described by the Langevin dynamics,~\cite{minnhagen-rev,ahns} 
we can view the
2D Coulomb gas model with Langevin dynamics as the dynamic model for the ``pure''
2D vortex response which includes neither any coupling to a spin-wave 
part nor any pinning. 
For a superconducting film in the absence of an external
perpendicular magnetic field (which corresponds to the absence of a net
rotation in case of superfluid $^4$He film), the numbers of vortices
and antivortices (with vorticities 1 and $-1$, respectively)
are always equal, and the system 
is described by the charge neutral two components 2D Coulomb 
gas model,~\cite{minnhagen-rev} on which the AHNS,
as well as the MP, are based.
Although it was clearly 
shown from simulation of this neutral 2D Coulomb gas model that the
low-temperature phase is well described by the MP response form with the
logarithmic divergence of the conductivity,~\cite{holmlund} 
there still remains the possibility that the coupling to the spin-wave
part or pinning could also influence and perhaps lead to the anomalous
response in certain cases.

In Ref.~\onlinecite{theron} it has been found in experiments that 
the 2D triangular Josephson array in a perpendicular
magnetic field  is also well described by the MP response form,
which has again been confirmed in simulations of the related frustrated
2D $XY$ model with relaxation dynamics.~\cite{jonsson1,jonsson2}
Since in these cases pinning could with large probability be excluded
as the cause of the effect, the observed anomalous behavior can in these cases
be attributed to either the logarithmic vortex interaction, or to the interplay
between the vortex and the spin-wave interactions, or to a combination of 
both.~\cite{theron,jonsson2,beck,korshunov,capezalli}
From the analysis of the experimental results in 
Ref.~\onlinecite{kapitulnik} for thin MoGe 
superconducting film in a perpendicular magnetic field, 
it has also been pointed out that the data show MP behavior.~\cite{jonsson1}
Since in Ref.~\onlinecite{kapitulnik} it was argued that
for this sample the vortex pinning due to disorder
was important, this raises the question of how pinning influences 
the dynamic behavior.
Could the MP response form also arise from an interplay between the
vortex system and pinning? In the present work we conclude,
from simulations of the lattice Coulomb gas model,
that the vortex system, in the absence of couplings to the
spin-wave part but in the presence of pinning, indeed in certain parameter
regions gives rise to MP features reflecting an anomalous diffusion.

The paper is organized as follows:
In Sec.~\ref{sec:lcg} we describe the Coulomb gas model and the correlation
functions we study in simulations. Section~\ref{sec:dynamic} gives a brief
description of the features of the MP response form and make comparisons
with the standard Drude response form.
The results from the simulations are given in Sec.~\ref{sec:results},
and in Sec.~\ref{sec:discussion} we try to give some perspective 
from comparisons with other results.     

\section{The Lattice Coulomb Gas Model} \label{sec:lcg}

In this work, we use the 2D Coulomb gas model on an $L \times L$
triangular lattice, whose Hamiltonian in the absence of disorder 
is written as~\cite{lcg}
\begin{equation} \label{eq:H0}
H = \frac{1}{2} \sum_{ij} (n_i - f) V_{ij} (n_j - f), 
\end{equation}
where $n_i$  is the integer charge on the $i$th site and 
the frustration
$f$ controls the total number of charges $N_c \equiv \sum_i n_i 
= f N$ ($N \equiv L^2$).
The interaction $V_{ij}$ between charges at positions
$\bfr_i$ and $\bfr_j$ for a triangular lattice is given by
\begin{equation}
V_{ij} = \frac{1}{N}\sum_{\bfk \neq 0 } 
\frac{3\pi e^{i\bfk\cdot(\bfr_i -\bfr_j)}}
{ 6 - 2\cos\bfk\cdot{\bf a}_1 - 2\cos\bfk\cdot{\bf a}_2
- 2\cos\bfk\cdot{\bf a}_3}, 
\end{equation}
where $\bfa_1 \equiv \bfa$, $\bfa_2 \equiv \bfb$, 
and $\bfa_3 \equiv \bfa - \bfb$ with two primitive translational 
vectors $\bfa$ and $\bfb$ for 2D triangular lattice.
(We choose $\bfa = \hat\bfx$ and $\bfb = \hat\bfx/2 + \sqrt{3}\hat\bfy/2$
with lattice spacing set to unity.)

We also study the effects of quenched disorder and consider 
the Hamiltonian 
\begin{equation} \label{eq:H}
H = \frac{1}{2} \sum_{ij} (n_i - f) V_{ij} (n_j - f)
  - \sum_i U_i n_i , 
\end{equation}
where the pinning potential $U_i$ at $i$ has value $U^p (>0)$ if the 
disorder is located at $i$
and $U_i = 0$ otherwise. The randomly distributed
disorder realization is parameterized by the pinning strength $U^p$
and the ratio $r_p$ between the number of pinning sites $N_p$ and the 
total number of charges $N_c$: $r_p \equiv N_p / N_c$. This
corresponds to point disorder since we can express it as
$U(\bfr)  = \sum_{j=1}^{N_p} U^p \delta(\bfr -  \bfr_j) $
in continuum limit, where $\bfr_j$ denotes the positions of disordered sites.

We use Monte Carlo (MC) dynamics which can be implemented as follows: 
Allow two different MC tries where one is the 
charge hopping to one of its nearest neighbors, 
i.e., $(1,0) \rightarrow (0,1)$, 
and the other is the generation of a charge-anticharge dipole on 
one bond, i.e., $(0,0) \rightarrow (1,-1)$. 
As was already observed in Ref.~\onlinecite{lcg},
for parameters used in this work the acceptance ratio
for generation of charge dipole is found to be extremely small 
even at temperatures much higher than the critical temperatures
(referring to the depinning and melting of the solid phase).
We are here studying a temperature region where no charge dipoles are in
practice created.
Consequently, 
we only include the charge hopping in our MC try. This means that the
model we are studying only contain charges of one sign. Our algorithm
is thus:
At each step, we first pick up one charge at random
and try to move it to one of its six nearest neighboring sites 
which is also randomly chosen. This MC try is accepted or not 
according to the standard Metropolis algorithm.
One can make the MC simulation more efficient by using a
nonlocal update (e.g., the randomly chosen charge can be allowed to 
hop to a distant site). On the other hand, as far as the dynamic behavior
is concerned, the local update rule like the one used in the present work
is believed to better imitate the relaxational dynamics of the 
corresponding continuum model (which corresponds to the Coulomb gas model
with Langevin dynamics).~\cite{nowak}

For the characterizations of the thermodynamic properties,
we measure (i) the dielectric constant $1/\epsilon(0)$, 
which corresponds to the
helicity modulus in the vortex systems and detects the superconducting to normal
transition, (ii) the orientational order parameter $\phi_6$, 
which probes the melting of the solid-like structure, and (iii) the 
specific heat $C_v$ (see Ref.~\onlinecite{lcg}):
\begin{eqnarray}
\frac{1}{\epsilon(0)} &\equiv& \lim_{\bfk \rightarrow 0}
\left[ 1 - \frac{2\pi}{TNk^2}\langle n_\bfk n_{-\bfk} \rangle \right] , 
\label{eq:eps} \\
\phi_6 &=& \frac{1}{N_c^2}\sum_{ij}\langle e^{i6(\theta_i -\theta_j)}\rangle, \\
C_v &=& \frac{\langle H^2 \rangle - \langle H \rangle^2}{T^2 N}, 
\end{eqnarray}
where $\langle \cdots \rangle$ is the thermal average, $n_\bfk \equiv 
\sum_i n_i e^{-i\bfk\cdot\bfr_i}$ is the Fourier transformation of the 
charge distribution, $\theta_i$ is the angle between a reference direction, 
say $x$, and the line connecting the $i$th particle and its closest neighbor.
Since the smallest wavevector is limited by the system size, 
we choose $\bfk = \bfA/L$, $\bfB/L$, and $-(\bfA + \bfB)/L$ with
reciprocal lattice vectors $\bfA \equiv 4\pi\hat\bfy/\sqrt{3}$
and $\bfB \equiv 2\pi(\hat\bfx -\hat\bfy/\sqrt{3})$, and take
the average over these three smallest $\bfk$'s (with magnitude $k = \sqrt{16\pi^2/3N}$)
to obtain the static dielectric
function $1/\epsilon(0)$ in Eq.~(\ref{eq:eps}).

The dynamic behaviors are described by the dynamic dielectric
function $1/\epsilon(\omega)$
(see Refs.~\onlinecite{houlrik} and \onlinecite{holmlund} for details):
\begin{eqnarray}
& &  \frac{1}{\epsilon(\omega )}-\frac{1}{\epsilon(0)}=-i\omega\int_0^\infty dte^{i\omega t}G(t), \label{eq:epsomega} \\
& &G(t)\equiv \lim_{k\rightarrow 0}\frac{2\pi}{Nk^2T}\langle n_\bfk(t) n_{-\bfk} (0) \rangle ,   \label{eq:g}
\end{eqnarray}
where $\langle n_\bfk (t) n_{-\bfk} (0) \rangle$ is the charge
correlation function, and $1/\epsilon(0)$ is the static
dielectric constant given above.
Just as for $1/\epsilon(0)$ in Eq.~(\ref{eq:eps}), $G(t)$ is obtained as the average
over the three smallest wavevectors.
We measure time $t$ in units of the
MC step so that one time unit corresponds to $N_c$ MC tries to move the
randomly chosen particle to one of its nearest neighbors.

\section{Dynamic Response} \label{sec:dynamic}
The dynamic response function we focus on is the complex dielectric function
$1/\epsilon(\omega )$ given by Eq.~(\ref{eq:epsomega}).
 The anomalous vortex
 diffusion is reflected in the fact $G(t)\propto 1/t$ for large $t$.~\cite{minnhagen-exp}
 From this one directly infers that to leading order in small
 $\omega$ (Ref.~\onlinecite{beom-big}) 
\begin{eqnarray} 
& & {\rm Re}\left(\frac{1}{\epsilon (\omega)}-\frac{1}{\epsilon (0)}\right)=
\frac{1}{\tilde{\epsilon}} \omega   , \label{eq_reps} \\ 
& &
{\rm Im}\left(\frac{1}{\epsilon(\omega)}-\frac{1}{\epsilon(0)}\right)= {\rm
Im}\left(\frac{1}{\epsilon(\omega)}\right)=\frac{2}{\tilde{\epsilon}\pi}\omega
\ln \omega  ,\label{eq_ieps} 
\end{eqnarray}
where $1/\tilde{\epsilon}$ is a constant and
the first equality in Eq.~(\ref{eq_ieps}) follows because
$1/\epsilon(0)$ is always a real quantity.
The complex conductivity for a superconductor $\sigma(\omega)$ corresponds 
to $-1/i\omega \epsilon(\omega)$
which means that ${\rm Re}\sigma(\omega )\propto -\ln \omega$ for small
$\omega$.~\cite{minnhagen-rev}
This logarithmic divergence is another characteristics of the anomalous vortex diffusion.

The MP form is given by
\begin{eqnarray} 
  {\rm Re}\left(\frac{1}{\epsilon(\omega)}-\frac{1}{\epsilon(0)}\right)=
    \frac{1}{\tilde{\epsilon}}\frac{\omega}{\omega+\omega_0} , \label{eq_rmp} \\
  {\rm Im}\left(\frac{1}{\epsilon(\omega)}\right)=
  -\frac{2}{\tilde{\epsilon}\pi}\frac{\omega \omega_0 \ln
    \omega/\omega_0}{\omega^2-\omega_0^2}  , \label{eq_imp}
\end{eqnarray}
and the corresponding explicit form of $G(t)$, which we denote by $G_{{\rm
MP}}$, hence incorporates the features associated with
$G(t)\propto 1/t$.~\cite{beom-big} The MP form is associated with a
single characteristic frequency $\omega_0$.  In the 2D neutral Coulomb gas
all vortices are bound in vortex-antivortex pairs in the
low-temperature phase.~\cite{minnhagen-rev,kosterlitz} This implies an
infinite correlation length which for a superconductor means a
vanishing resistance $R$.~\cite{minnhagen-rev} Since the logarithmic
divergence of the conductivity means that $R=0$, a possible scenario is that
the MP response form describes the response of the
vortex-antivortex bound pairs and hence should apply to
the low-temperature phase. It has been verified, from
simulations of the 2D neutral Coulomb gas with Langevin dynamics,
that the MP form indeed gives a very good description in this
case.~\cite{holmlund} Above the KT transition there are both free
vortices and bound vortex-antivortex pairs present and hence there
is a finite correlation length. This means that $G(t)$ has an
exponential decay of the form $e^{-t/\tau}$, where $\tau$ is a
relaxation time and $R\propto 1/\tau$. Thus a finite resistance is
associated with an exponential decay of $G(t)$. If there were only
free vortices then $G(t)$ would be dominated by the exponential factor
and the response should to good approximation be described by
the standard Drude form $G_{\rm D}\propto e^{-t/\tau}$, i.e.,
\begin{eqnarray} 
& & {\rm Re}\left(\frac{1}{\epsilon(\omega)}\right)=
  \frac{1}{\tilde{\epsilon}}\frac{\omega^2}{\omega^2+\tau^{-2}} , 
\label{eq_rd} \\
& &  {\rm Im}\left(\frac{1}{\epsilon(\omega)}\right)=
  -\frac{1}{\tilde{\epsilon}}\frac{\omega\tau^{-1}}{\omega^2+\tau^{-2}}  .
\label{eq_id}
\end{eqnarray}
More generally, when free vortices are present the response form
$G_{\rm MPD}=G_{\rm MP}(t)G_{\rm D}(t)$ has from simulations been shown to give
  a good parameterization of the data.~\cite{holmlund,beom-big,beom-flux}

One difference between the MP response and the Drude response, which
is easy to gauge in experiments and simulations, is the peak
ratio: The peak ratio (PR) is defined as the ratio between the
real and
imaginary part of the complex dielectric function at the frequency where
$|{\rm Im}[1/\epsilon(\omega)]|$ has its maximum,
i.e., at the dissipation peak (see Fig. 6). For the standard Drude response this
ratio is 1 whereas it is $2/\pi\approx 0.64$ for the MP response.
In general when the resistance is finite the peak ratio should be between
these two values, $2/\pi\leq {\rm PR}\leq 1$. 

Our strategy is to calculate $G(t)$ as described above and then to
gauge the response by studying the peak ratio, $G(t)$ for large $t$,
and to what extent the response is parameterized by the MP form.

\section{Simulation Results} \label{sec:results}
\subsection{Lattice Coulomb gas without disorder}
\label{subsec:pure}
In this section, we first present the results of the MC simulations
of the lattice Coulomb gas model given by the Hamiltonian
in Eq.~(\ref{eq:H0}) for the case without random pinning sites, i.e., $U_i=0$
in Eq.~(\ref{eq:H}). 
It is known that the system can undergo two phase
transitions in this case:~\cite{lcg} One is the depinning of the solid
at $T_c$,
corresponding to the superconducting to normal transition,
and the other is the solid to liquid transition at $T_m$ 
($T_c\leq T_m$). These two transitions can be identified by aid of the
quantities
$1/\epsilon(0)$ and $\phi_6$, respectively.
Since the continuum system corresponds to 
an infinitesimally small lattice spacing and the frustration $f$ is 
proportional to the area of the elementary plaquette, 
the continuum limit of the lattice Coulomb gas is given by
$f \rightarrow 0$, but with the number of particles $N_c$ still finite:
$N_c = N f$=const.
The observation in Ref.~\onlinecite{lcg} that $T_c$ becomes
smaller as $f$ is decreased 
indicates that $T_c \rightarrow 0$ in the continuum system
so that the system only has one transition at a nonzero temperature $T_m$.

\begin{figure}
\centering{\resizebox*{!}{6.0cm}{\includegraphics{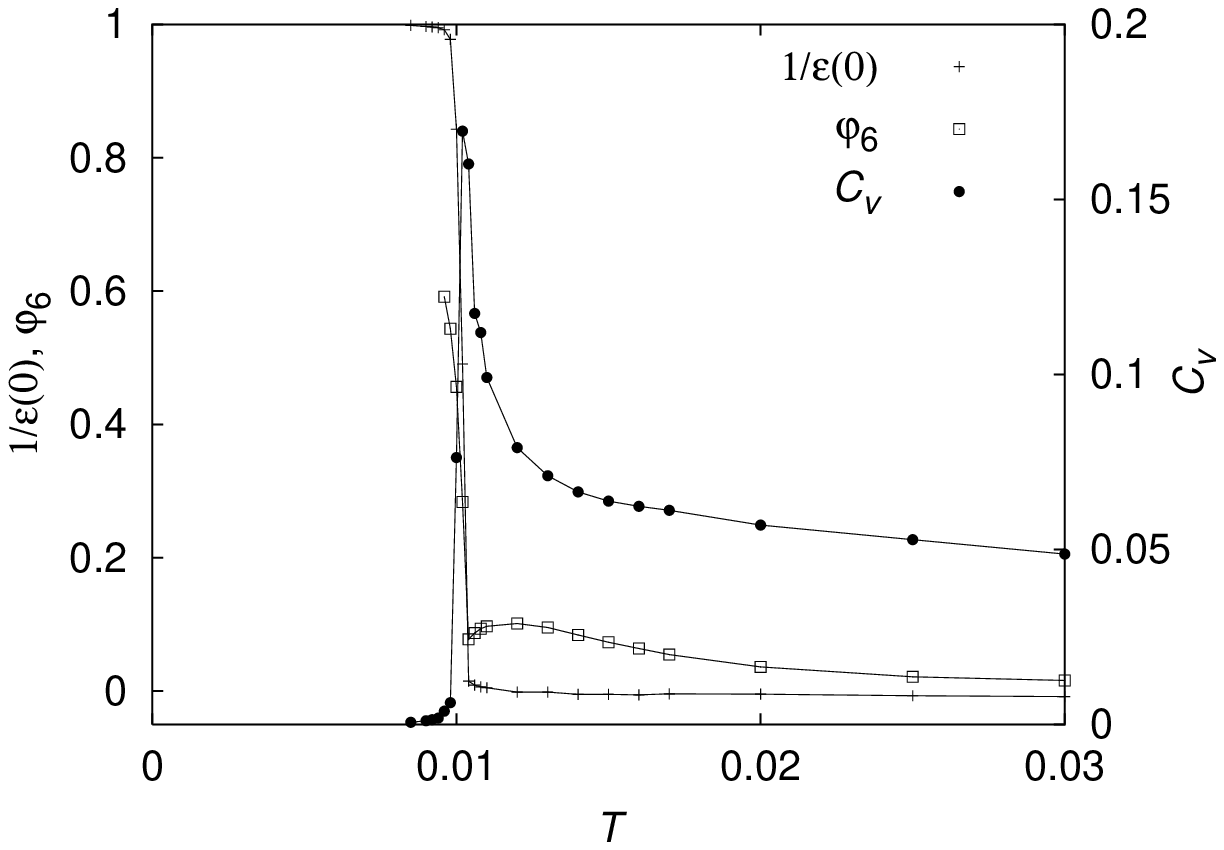}}}
\caption{
Static dielectric function $1/\epsilon(\omega = 0)$, 
the orientational order parameter $\phi_6$, and the
specific heat $C_v$ vs temperature $T$
for the 2D lattice Coulomb gas model on a $64\times 64$ triangular 
grid with the frustration $f = 1/16$.
The depinning transition at $T_c$, corresponding to
the superconducting to normal transition, is signaled by an
abrupt change of $1/\epsilon(0)$.
The solid to liquid transition at $T_m$ probed by $\phi_6$  
occurs at approximately
the same temperature, i.e., $T_c \approx T_m \approx 0.01$.
$C_v$ also shows a sharp maximum near the same temperature.
}
\label{fig:static16}
\end{figure}

We start by presenting some static results for $f=1/16$ on
a $64\times 64$ triangular grid. In the simulations, we start from high enough
temperatures and anneal the system by decreasing the temperature
slowly, to avoid being captured by metastable states at low
temperatures. However, the local MC scheme used in this work
(a particle is only allowed to hop to its nearest neighbor sites) makes it
difficult to achieve equilibration near and below the transition.
Thus very many MC updates are required in this region (our longest
runs consists of $2\times 10^6$ MC steps).
As $T$ is decreased $1/\epsilon(0)$ abruptly changes to unity
near $T \approx 0.01$ (see Fig.~\ref{fig:static16}). This abrupt change 
is associated with the normal (high $T$) to superconducting (low $T$) 
transition in 
the vortex system since $1/\epsilon(0)$ corresponds to the superfluid
density.~\cite{minnhagen-rev} In the Coulomb gas model it corresponds
to a transition where the Coulomb gas systems becomes pinned to the underlying grid.
The six-fold orientational order parameter $\phi_6$ has the value
unity for a perfect triangular Abrikosov vortex lattice. Consequently the
abrupt drop of $\phi_6$ seen
in Fig.~\ref{fig:static16} indicates that the vortex solid melts into a
vortex liquid near $T \approx 0.01$. 
The specific heat $C_v$ also shows a sharp maximum
near this temperature $T \approx 0.01$. Consequently the data in
Fig.~\ref{fig:static16} for the lattice
Coulomb gas model with $f=1/16$ are consistent with a single phase transition
corresponding to a transition where the superconductor becomes normal
and the Abrikosov vortex structure melts, i.e., $T_c \approx T_m \approx 0.01$.~\cite{lcg}

\begin{figure}
\centering{\resizebox*{!}{6.0cm}{\includegraphics{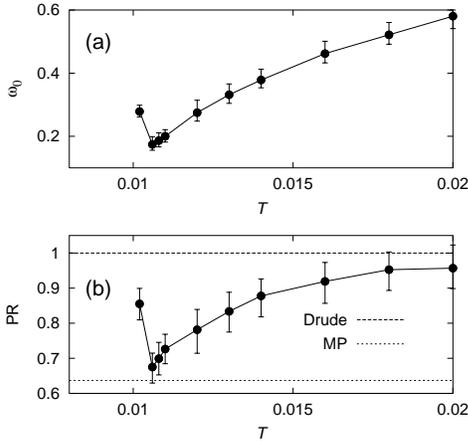}}}
\caption{
Characteristic features of the dynamic dielectric function 
for the system with the frustration $f = 1/16$ and size $L = 64$. (a) The characteristic frequency
scale $\omega_0$ as a function of $T$ obtained from the peak position of 
$|{\rm Im}[1/\epsilon(\omega)]|$: $\omega_0$
decreases near $T \approx 0.01$, reflecting a critical slowing down,
in accordance with Fig.~\ref{fig:static16}.
(b) The peak ratio PR defined by the ratio between real and imaginary
part of $1/\epsilon(\omega)$: The response function is described by the 
simple Drude response form with PR$=1$ at high enough temperatures 
and crosses over to the anomalous behavior
with PR$=2/\pi$ as $T$ is decreased.~\cite{foot3}
}
\label{fig:dynamic16}
\end{figure}

We next turn to the dynamic results for $f=1/16$.
These are shown in Fig.~\ref{fig:dynamic16}:
The characteristic frequency $\omega_0$ in 
Fig.~\ref{fig:dynamic16}(a), determined from the peak
position of $|{\rm Im}[1/\epsilon(\omega)]|$, shows a dip-like feature
near $T = 0.01$, reflecting a critical slowing down close to the
phase transition (compare the static results in Fig.~\ref{fig:static16}).
In general, a system composed of particles 
is expected to be well described by a Drude approximation at sufficiently high
temperatures because the interaction between the particles becomes
negligible in comparison with the strong thermal fluctuations.
The dynamic response function in such a temperature regime
should be well described by the Drude response form 
Eqs.~(\ref{eq_reps}) and (\ref{eq_ieps}) and should consequently be well
characterized by the peak ratio (PR) equal to one.
As is shown in Fig.~\ref{fig:static16}(b), the dynamic dielectric
function indeed approaches the Drude PR value one at higher temperatures. However, the PR  crosses over 
to a value close to PR=0.64 which characterizes the anomalous behavior
(see the previous section) as we approach $T \approx 0.01$ from above.~\cite{foot1}

The question is then what causes this crossover behavior. Is it
associated with the depinning transition
at $T_c$ or with the melting transition at $T_m$? In order to address
this question one needs a temperature separation between the two transitions.
In Ref.~\onlinecite{lcg}, it was shown that as $f$ is decreased,
the two transitions present become well separated.
With this in mind we, as a next step, repeat the simulations for a much smaller value of $f$.

\begin{figure}
\centering{\resizebox*{!}{5.5cm}{\includegraphics{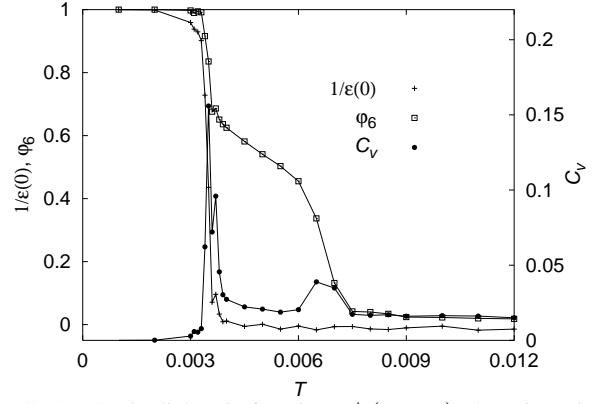}}}
\caption{
Static dielectric function $1/\epsilon(\omega = 0)$, 
the orientational order parameter $\phi_6$, and the
specific heat $C_v$ vs temperature $T$
for the system with $f=1/64$ and $L=96$.
The depinning transition is 
at $T_c \approx 0.0035$ and melting
transition is at $T_m \approx 0.007$. Also note that
$C_v$ displays two peaks: one near $T_c$ and the other
one near $T_m$. The intermediate phase existing between $T_c$
and $T_m$ has a nonzero orientational order and corresponds to
a floating solid phase.
}
\label{fig:static64}
\end{figure}

Figure~\ref{fig:static64} shows the static results $1/\epsilon(0)$,
$\phi_6$, and $C_v$ for the system with $f=1/64$ on a $96\times 96$ lattice.
As seen from the figure, the superconducting to normal transition is
now at 
$T_c \approx 0.0035$ and is well separated from the vortex solid melting transition at 
$T_m \approx 0.007$, as was found in Ref.~\onlinecite{lcg}. 
As also seen in Fig~\ref{fig:static64}, the existence of two
transitions are also reflected by two peaks in $C_v$: 
one near $T_c$ and another broader one near $T_m$. 
The intermediate phase existing between
$T_c$ and $T_m$ is characterized by a vanishing $1/\epsilon(0)$ together
with a nonzero $\phi_6$ and is interpreted as a floating solid phase:~\cite{lcg}
The vortex solid structure is depinned from the underlying discrete
grid and floats around, which leads to dissipation of energy. 

\begin{figure}
\centering{\resizebox*{!}{6.0cm}{\includegraphics{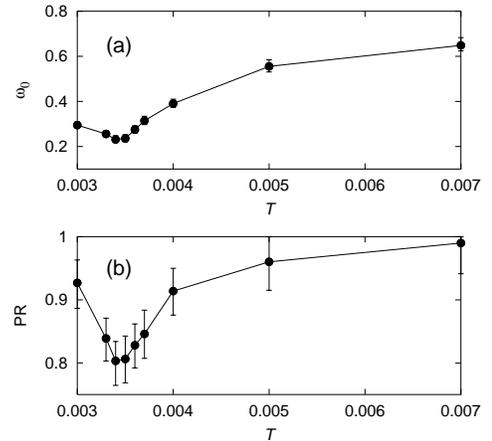}}}
\caption{
Characterization of the dynamic dielectric function 
for the system with the frustration $f = 1/64$ and size $L = 96$. 
(a) The characteristic frequency $\omega_0$ vs temperature $T$
shows a dip-like structure near $T \approx T_c \approx 0.0035$
(see Fig.~\ref{fig:static64} for the static results).
(b) The peak ratio PR vs $T$ reflects a crossover behavior from
Drude response form in high temperatures to the anomalous one
near $T_c$.~\cite{foot3}
}
\label{fig:dynamic64}
\end{figure}

Next we analyze the dynamic response function for $f=1/64$ is terms 
of the characteristic frequency $\omega_0$ and the peak ratio PR. 
In Fig.~\ref{fig:dynamic64}, it is shown that $\omega_0$ displays
a dip-like structure, reflecting a critical slowing down near $T_c$
rather than at $T_m$.~\cite{note1} Similarly, the correspondence of the crossover behavior for $f=1/16$
in PR from Drude to anomalous response form shown in
Fig.~\ref{fig:dynamic16}(b),
is in  Fig.~\ref{fig:dynamic64}(b) found to be near $T_c$ and not near
$T_m$.~\cite{note1} From this we conclude that the crossover behavior in the
dynamic response is associated with $T_c$ and hence with the pinning
of the Coulomb gas to the underlying lattice. Thus for the pure one
component lattice Coulomb gas with MC dynamics we infer that there is a crossover
behavior from a normal Drude like response to an anomalous response.
An essential ingredient for the appearance of the anomalous response is associated with the pinning
between the vortex system and the underlying lattice.

\subsection{Lattice Coulomb gas with disorder}
\label{subsec:disorder}

In the previous section we inferred that the pinning, caused by the
underlying lattice, leads to a crossover from a Drude to an anomalous dynamic
response. In the present section we study the role of pinning further
by introducing extrinsic vortex pinning caused by disorder into the
model. The general idea is that the anomaly in the dynamics is caused
by a sluggish motion of the vortex system and that pinning hindering
the vortex motion can cause such a behavior. To this end we introduce
point impurities into the lattice Coulomb gas  model
as described by the Hamiltonian in Eq.~(\ref{eq:H}). For a
superconductor this corresponds to magnetic point impurities.
In this section, we perform MC dynamic simulations for
10 different disorder realizations in order to get
disorder averaged quantities. 

\begin{figure}
\centering{\resizebox*{!}{5.5cm}{\includegraphics{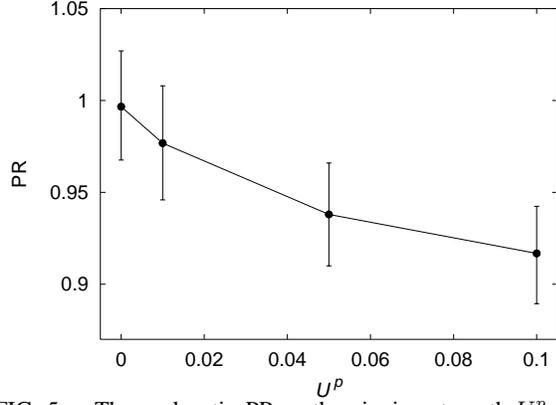}}}
\caption{
The peak ratio PR vs the pinning strength $U^p$ for the system
with $f=1/64$ and $L=96$ at $T=0.01$. The number of pinning sites
was fixed to 14 (which is about 10\% of the total number of vortices
and the results correspond to an average over 10 random pinning realizations).
The PR starts from unity for the pure case ($U^p = 0$), which is the value
for the Drude response, and decreases as $U^p$ is increased, reflecting
that the vortex motion becomes more sluggish. 
}
\label{fig:prpin}
\end{figure}

In Fig.~\ref{fig:prpin} the PR for the system with $f=1/64$ and $L=96$
at $T=0.01$ is displayed as a function of the pinning strength $U^p$ (the
number of pinning sites $N_p$ is fixed to 14, which approximately corresponds
to 10\% of the total number of vortices $N_c$, i.e., $N_p \approx 0.1
N_c$).
As seen in Fig.~\ref{fig:prpin} the dynamic response
for the pure case ($U^p = 0$) has PR close to one at this higher $T$
[compare Fig.~\ref{fig:dynamic64}(b)]. This means that the pinning to
the underlying lattice at this $T$ is too weak to influence the
behavior so that the vortices should obey a Drude response.
However, if we introduce disorder then Fig.~\ref{fig:prpin} shows a
decreasing PR with increasing pinning strength again suggesting a crossover
towards an anomalous dynamic response [compare Fig.~\ref{fig:dynamic64}(b)]. 
This is consistent with the idea that when 
the pinning strength becomes stronger, the motion of the vortices becomes more
sluggish, resulting in a  decrease of the PR and a crossover to an
anomalous response.

\begin{figure}
\centering{\resizebox*{!}{6.0cm}{\includegraphics{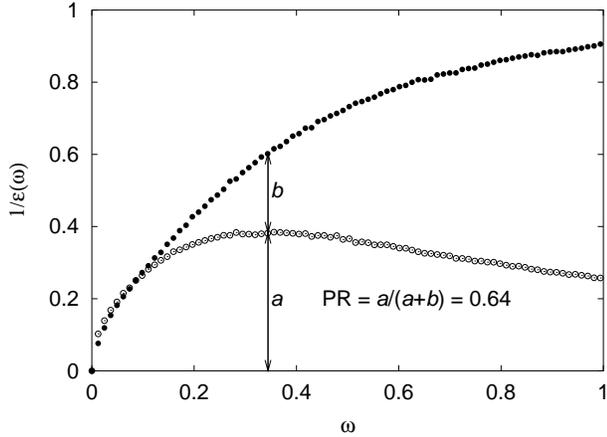}}}
\caption{
The dynamic dielectric function obtained 
for the system with $f=1/64$ and
$L=96$ in the presence of extrinsic pinning with
$U^p = 0.1$ and $N_p = N_c$ for $T=0.02$ (10 random pinning
configurations were used). Filled circles and open circles
correspond to ${\rm Re}[1/\epsilon(\omega)]$ and  $|{\rm
  Im}[1/\epsilon(\omega)]|$. 
The PR (the ratio between the real and imaginary part at the 
maximum of the imaginary part) is found to be close
to the MP value $2/\pi$.
}
\label{fig:ew}
\end{figure}

\begin{figure}
\centering{\resizebox*{!}{6.0cm}{\includegraphics{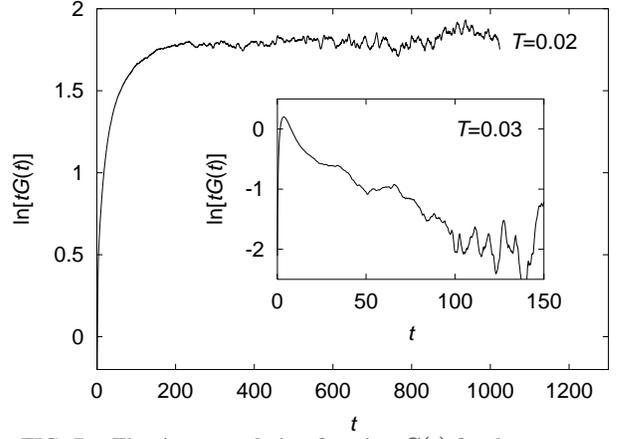}}}
\caption{
  The time-correlation function $G(t)$ for the same case as in Fig.~\ref{fig:ew}
 plotted as $\ln t G(t)$
 vs $t$. The horizontal plateau for larger $t$ shows that the 
 long-time behavior of $G(t)$ has a $1/t$ decay. The inset shows the
 same case at $T=0.03$ and at this higher temperature $G(t)$ appears
 to have an exponential decay.
}
\label{fig:Gt}
\end{figure}

\begin{figure}
\centering{\resizebox*{!}{6.0cm}{\includegraphics{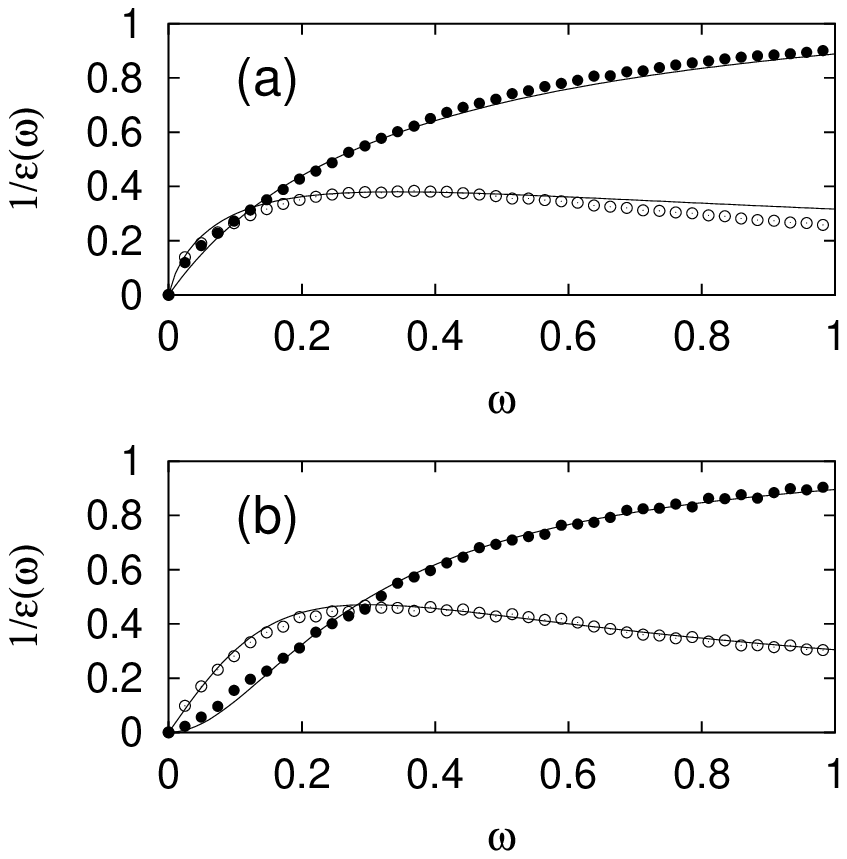}}}
\caption{Fits of the dynamic dielectric function to the MP form.
  (a) $1/\epsilon(\omega )$ corresponding to
  the anomalous $G(t)$ for $T=0.02$ in Fig.~\ref{fig:Gt} is fitted to
  Eqs.~(\ref{eq_rmp}) and (\ref{eq_imp}). Filled circles, open circles and full
  drawn curves correspond to ${\rm Re}[1/\epsilon(\omega)]$, $|{\rm
  Im}[1/\epsilon(\omega)]|$, and the MP form, respectively. 
  A reasonably good fit is obtained.
  (b) $1/\epsilon(\omega )$, corresponding to
  the exponentially decaying $G(t)$ for $T=0.03$ in the inset in
  Fig.~\ref{fig:Gt}, is fitted to the parameterization obtained from
  $G_{\rm MPD}(t) \equiv G_{\rm MP}(t)G_{\rm D}(t)$ (see the text). Filled circles, open circles and full
  drawn curves correspond to ${\rm Re}[1/\epsilon(\omega)]$, $|{\rm
  Im}1/\epsilon(\omega)|$, and the $1/\epsilon(\omega)$ from
  $G_{\rm MPD}$ , respectively. A reasonably good fit is
  found for PR$\approx0.9$ which is slightly smaller than PR$=1$ 
  corresponding to the pure Drude value. 
}
\label{fig:ewall}
\end{figure}

This scenario implies that if we introduce enough pinning then the
crossover to the anomalous response should become complete. 
To achieve this we investigate a stronger pinning case where $N_p = N_c$ and the pinning
strength $U^p = 0.1$. Figure~\ref{fig:ew} shows the obtained dielectric
function $1/\epsilon(\omega)$ for this case [see
Eqs.~(\ref{eq:epsomega}) and (\ref{eq:g})] at $T=0.02$ which is much higher than $T_c$ as well as $T_m$ for
the pure case ($T_c \approx 0.0035$ and $T_m \approx 0.007$,
respectively). As illustrated in Fig.~\ref{fig:ew},
the PR is indeed close to the anomalous value PR$=2/\pi$ in this
case. In the MP phenomenology the anomalous dynamics is
linked to a $1/t$ decay of the correlation function $G(t)$ [compare
Eqs.~(\ref{eq:g})-(\ref{eq_imp})].  
The time-correlation function $G(t)$ (corresponding to
$1/\epsilon(\omega)$ in Fig.~\ref{fig:ew}) is shown in Fig.~\ref{fig:Gt}
and indeed has a $1/t$-tail as is
manifested by the horizontal plateau of $tG(t)$ for larger $t$. This suggests that there exists a link between a $1/t$-tail in
$G(t)$ and the PR$=2/\pi$. Such a link is incorporated into the MP response form Eqs.~(\ref{eq_rmp}) and (\ref{eq_imp}). The inset in Fig.~\ref{fig:Gt} shows what
happens for the same case at the higher $T=0.03$. In this case $G(t)$ suggests
the presence of an exponential decay. This is very similar to previous
simulations where $G_{{\rm MPD}}\equiv G_{{\rm MP}}(t)G_{\rm D}(t)$ has been
shown to give a good parameterization [see Sec.~\ref{sec:dynamic} below
Eq.~(\ref{eq_id}) and Refs.~\onlinecite{holmlund}, \onlinecite{beom-big}, and
\onlinecite{beom-flux}].  Figure~\ref{fig:ewall}(a) shows that a reasonable fit
of $1/\epsilon(\omega)$ to the MP form is obtained in the anomalous case
$T=0.02$. This suggests that the basic link between the $G(t)\propto 1/t$ and
the PR=$2/\pi$ is reasonably well
captured by the MP form.  
Figure~\ref{fig:ewall}(b) shows a fit to the exponentially decaying case
at $T=0.03$ using the parameterization $G_{\rm MPD} \equiv G_{\rm MP}(t)
G_{\rm D}(t)$. Again a
reasonable fit is obtained with PR$\approx0.9$. This suggests that,
although it is close to the Drude form, there is a small crossover towards the
anomalous response (see Fig. \ref{fig:prpin}).

In this section we have thus shown that for the one component lattice
Coulomb gas, at a $T$ so high that the pinning to the underlying
lattice plays no role, it is still possible to obtain a crossover to
the anomalous response by introducing pinning through randomly
distributed pinning sites.  

\section{Comparison and Discussions} \label{sec:discussion}

We have found from simulations that the 2D one component lattice
Coulomb gas with Monte Carlo dynamics under certain circumstances
displays an anomalous dynamic response. An essential ingredient for
the appearance of this anomalous response is found to be pinning;
either pinning due to the underlying lattice or external pinning
induced by disorder. The anomalous response vanishes gradually as the pinning effects become weaker and from this we conclude that there is no anomaly in the absence of pinning. 

We gauge the degree
of anomaly in the response by the behavior of the correlation function
$G(t)$ for large $t$, the peak ratio, and to what extent the response
can be parameterized by the MP form. We find that the full anomaly is
characterized by $G(t)\propto 1/t$ for large $t$, a peak ratio
consistent with $2/\pi$ and note that these two features are also
incorporated into the phenomenological MP form. For the
one component lattice Coulomb gas this means that an anomalous
response in the absence of externally introduced pinning only appears
very close to the depinning of the vortex system from the underlying
lattice structure. From this we further infer that  
the pure continuum one component Coulomb gas has no anomalous
response and that in this case external
pinning is required to change the response from normal to anomalous.

It is interesting to compare this with the 2D neutral (two components)
continuum Coulomb gas with Langevin dynamics which is the generic
model
for the vortex physics of a 2D superconductor(superfluid) in the
absence of an external perpendicular magnetic field (net
rotation).~\cite{minnhagen-rev} This model undergoes a KT transition
and below this transition the dynamic response has from simulations
been shown to
be anomalous characterized by $G(t)\propto 1/t$ for large $t$, a peak ratio
consistent with $2/\pi$, and to be well described by the MP form.~\cite{holmlund}    
Since the same response form is found for the one component Coulomb
gas in the presence of external pinning, as for the 2D neutral (two
component)
Coulomb gas below the KT transition, one might speculate about a common physical feature.
We here suggest that the common physical feature is pinning;
in the one case externally introduced and in the other intrinsic in
the sense
that negative vortices act as (moving) pinning centers for positive ones (and
vice versa).
In both cases the pinning mechanism suppresses the free vortex
diffusion.
When this suppression is complete the response becomes anomalous as
for the one component Coulomb gas with enough external pinning and for
the neutral Coulomb
gas below the KT-transition where all positive and negative vortices are bound together in pairs. 

Thus our conclusion based on simulations is that the 2D one component
{\em continuum} Coulomb gas cannot have an anomalous response by itself.
This is in contrast to the theoretical considerations in
Ref.~\onlinecite{capezalli},
based on a Mori approximation scheme, which suggest that the
one component model
in certain parameter regions could have an anomalous response. We have
not been able
to find any support for this suggestion in our simulations.

The MP form~\cite{minnhagen-rev} was originally motivated for the 2D
 neutral
 (two components) Coulomb gas as an extension of the AHNS
 phenomenology~\cite{ahns}
 and was assumed to describe the response of bound vortex
 pairs. Other
 recent improvements of the AHNS phenomenology also gives a peak ratio
 close
 to $2/\pi$.~\cite{bormann,bowley1} However, the MP form is to our
 knowledge the only
 form that incorporates both the large $t$ behavior $G(t)\propto 1/t$ 
 and the peak ratio $2/\pi$, both of which seems to be characteristics
 of the
 anomalous response. In the one component Coulomb gas case with
 external pinning
 there is at present really no motivation for the MP form; it is
 just
 a simple form which simultaneously incorporates two features which
 seem to be characteristics of the anomalous response.~\cite{foot2} 

 Our basic conclusion is that the anomalous response form can indeed
 arise
 from an interplay between pinning and a vortex system associated with
 an external perpendicular magnetic field. In this context one may
 note
 that a $2/\pi$ peak ratio was found for the experimental data in
 Ref.~\onlinecite{kapitulnik}.~\cite{jonsson1} This was a MoGe superconducting film 
 in a perpendicular magnetic field where pinning was expected to be
 important.~\cite{kapitulnik} We suggest that this might be an
 experimental
 example of our present findings from simulations.

\subsection*{Acknowledgment}
This work was supported by the Swedish Natural Research Council through contract No. FU 04040-332.

\newpage

\end{multicols}

\begin{thebibliography}{99}
\bibitem{foot1} In connection with Fig.~\ref{fig:Gt} for the 
disordered case the possibility of a link between PR=$2/\pi$
and $G(t)\sim 1/t$ is discussed. This link also seems to
be borne out in the present case without disorder although
the convergence of the data is not as impressive as in Fig.~\ref{fig:Gt}.
\bibitem{note1} The fact that $\omega_0$ does not reach zero in this
 case is due to the finite lattice size. We believe that this is also the reason why
  the PR does not decrease all the way to $2/\pi$.
\bibitem{foot2} The fact that disorder may cause a slow decay
of the correlations is reminiscent of glassy states. However,
the special feature of the anomalous response discussed here is that
the correlations decay precisely as $1/t$.
\bibitem{foot3} The cause of the increase of PR for decreasing $T$
at the lowest temperatures is not clear for us. It cannot be ruled
out that it may be an artifact of the finite-size lattice.
\end{thebibliography}
\end{document}